\documentclass[a4paper,11pt]{article}
\usepackage{a4wide,amsmath,amssymb,bbm}
\usepackage[english]{babel}

\newcommand{\tr}{\mathrm{tr}}
\newcommand{\Ad}{\mathrm{Ad}}
\makeatletter

\@addtoreset{equation}{section} \makeatother

\begin{document}

\hfill{ }

\vspace{30pt}

\begin{center}
{\huge{\bf Trivial solutions of generalized supergravity vs non-abelian T-duality anomaly}}

\vspace{80pt}

Linus Wulff

\vspace{15pt}

{\it\small Department of Theoretical Physics and Astrophysics, Masaryk University, 611 37 Brno, Czech Republic}\\

\vspace{100pt}

{\bf Abstract}
\end{center}
\noindent
The equations that follow from kappa symmetry of the type II Green-Schwarz string are a certain deformation, by a Killing vector field $K$, of the type II supergravity equations. We analyze under what conditions solutions of these `generalized' supergravity equations are trivial in the sense that they solve also the standard supergravity equations. We argue that for this to happen $K$ must be null and satisfy $dK=i_KH$ with $H=dB$ the NSNS three-form field strength. Non-trivial examples are provided by symmetric pp-wave solutions. We then analyze the consequences for non-abelian T-duality and the closely related homogenous Yang-Baxter sigma models. When one performs non-abelian T-duality of a string sigma model on a non-unimodular (sub)algebra one generates a non-vanishing $K$ proportional to the trace of the structure constants. This is expected to lead to an anomaly but we show that when $K$ satisfies the same conditions the anomaly in fact goes away leading to more possibilities for non-anomalous non-abelian T-duality.

\pagebreak 
\tableofcontents

\setcounter{page}{1}


\section{Introduction}
It was shown in \cite{Wulff:2016tju} that the equations for the target space fields which follow from the requirement of kappa symmetry of the type II Green-Schwarz superstring (or BRST invariance of the pure spinor string at the classical level\footnote{It was claimed in \cite{Berkovits:2001ue} that one gets the standard supergravity equations but extra assumptions, such as an $SL(2,\mathbbm R)$-invariant formulation in the IIB case, were made there. See also \cite{Mikhailov:2012id}.}) are in fact not, as previously thought, the standard type II supergravity equations but rather a certain deformation of these by a Killing vector $K$. When $K$ is set to zero the equations reduce to the standard supergravity equations. These generalized supergravity equations were first written down (in the bosonic sector) in \cite{Arutyunov:2015mqj} as the equations satisfied by the target space fields \cite{Arutyunov:2015qva} of the so-called $\eta$-deformed $AdS_5\times S^5$ superstring \cite{Delduc:2013qra}. They were interpreted as the conditions for one-loop scale invariance of the string sigma model, while the conditions for one-loop Weyl invariance are stronger, namely the standard supergravity equations. It was also shown in \cite{Arutyunov:2015mqj} that solutions of the generalized supergravity equations are related by T-duality, at the classical level in the sigma model ignoring in particular the shift of the dilaton, along the isometry defined by $K$ to solutions of standard supergravity.\footnote{When $K$ is time-like one gets a solution of type II* rather than type II supergravity. When $K$ is null one cannot carry out the T-duality directly. However, if there is a commuting null isometry, one can T-dualize in both null directions, which is equivalent to a time-like and a space-like T-duality, to get a solution of (type II*) supergravity, e.g. \cite{Orlando:2016qqu}.}

Here we will ask under what conditions it is possible for a solution to the generalized supergravity equations with $K\neq0$ to also be a solution of the standard supergravity equations (without doing T-duality). We will refer to such solutions as `trivial' since for these the deformation of the supergravity equations by $K$ becomes trivial. Naively it might seem that this should not be possible, however some explicit examples of such backgrounds are in fact known, a pp-wave example found in \cite{Hoare:2016hwh} (see also \cite{Roychowdhury:2018qsz}) and, very recently, certain deformations of $AdS_3\times S^3\times T^4$ in \cite{Sakamoto:2018krs}. These examples are closely related to so called homogeneous Yang-Baxter (YB) deformations of supercoset sigma models \cite{Klimcik:2002zj,Klimcik:2008eq,Kawaguchi:2014qwa}.

There is an interesting tension here with the standard expectation from sigma model anomalies. This tension comes about as follows. In was suggested in \cite{Hoare:2016wsk}, and demonstrated in several examples, that homogeneous YB deformations should be equivalent to non-abelian T-duality \cite{delaOssa:1992vci} on a centrally extended subalgebra (an important special case of this being so called TsT transformations \cite{Osten:2016dvf}).\footnote{Other, closely related, interpretations are e.g. \cite{Araujo:2017jkb,Sakamoto:2017cpu}.} This was subsequently proven in \cite{Borsato:2016pas} and a different perspective was also introduced -- YB models are equivalent to first adding a topological term (a closed $B$-field defined by a Lie algebra 2-cocycle) to the sigma model action and then performing non-abelian T-duality. The general construction of such deformed T-dual (DTD) models as applied to supercoset strings was worked out in \cite{Borsato:2017qsx}. It is known that the target space of YB sigma models in general solves the generalized supergravity equations (this follows since they can be formulated as Green-Schwarz strings) \cite{Borsato:2016ose}. The Killing vector $K$ turns out to be proportional to the trace of the structure constants of the subalgebra which was T-dualized on \cite{Borsato:2017qsx}. Therefore $K$ vanishes precisely when this subalgebra is what is known as unimodular. In fact, since the work of \cite{Alvarez:1994np,Elitzur:1994ri}, one expects a (mixed) anomaly for non-abelian T-duality on non-unimodular algebras.

But if it is possible to have solutions of generalized supergravity with non-vanishing $K$, which nevertheless also solve the standard supergravity equations, the corresponding sigma models should be Weyl invariant and the anomaly should be absent. Therefore there should be exceptions to the naive expectation that non-abelian T-duality on a non-unimodular algebra gives rise to an anomaly.

Here we will show that this is indeed the case and that the analysis based on the anomalous terms in the YB sigma model action agrees with an analysis based purely on the generalized supergravity equations. In particular we will argue that a trivial solution of the generalized supergravity equations should have $K$ null and satisfying $dK=i_KH$ with $H=dB$ the NSNS three-form field strength. Similarly we will show that these conditions are also precisely what is needed for the anomalous terms in the YB sigma model action to go away.

We will also show that any symmetric pp-wave solution of the generalized supergravity equations is of this form with $dK=i_KH=0$.

The outline of the rest of this note is as follows. First we recall the form of the generalized supergravity equations. We then discuss the simplest solutions, namely symmetric pp-wave spaces, and show that they typically have $K\neq0$ but also solve the standard supergravity equations. In section \ref{sec:trivial} we address the general question of when a solution to the generalized supergravity equations is trivial in the sense that it also solves the standard ones. The existence of such solutions is in tension with the expectations from non-abelian T-duality. We resolve this tension, in the context of bosonic YB models, in section \ref{sec:YB} by showing that in fact the anomalous terms go away in precisely these cases. We end with some conclusions.

\section{Generalized type II supergravity equations}
Here we will recall the generalized supergravity equations for the type IIB case. The type IIA equations can be written in an essentially identical form and everything we say will apply equally, modulo trivial replacements, to the type IIA case. The field content consists of the metric $g_{mn}$ and the NSNS two-form $B$, with field strength $H=dB$, just like in standard supergravity, but instead of the dilaton there is a one-form $X$ and the RR field strengths are replaced by $n$-form fields $\mathcal F^{(n)}$ which are no longer (a priory) defined in terms of potentials. In addition there is a (non-dynamical) Killing vector field $K$. These satisfy \cite{Arutyunov:2015mqj,Wulff:2016tju}\footnote{We set all fermionic fields to zero. Our conventions for differential forms are as follows. We write an $n$-form as $\alpha=\frac{1}{n!}e^{a_n}\wedge\cdots\wedge e^{a_1}\alpha_{a_1\cdots a_n}$ and the exterior derivative acts from the right. The inner product on forms is defined as
$\langle\alpha,\beta\rangle=\frac{1}{n!}\alpha_{a_1\cdots a_n}\beta^{a_1\cdots a_n}$ and the norm $|\alpha|^2=\langle\alpha,\alpha\rangle$. The Hodge dual is defined as
$$
*\alpha=\frac{1}{(10-n)!n!}e^{a_{10-n}}\wedge\cdots\wedge e^{a_1}\varepsilon_{a_1...a_{10-n}}{}^{b_n\cdots b_1}\alpha_{b_1\cdots b_n}\,,
$$
so that $*^2=1$ and $\alpha\wedge*\alpha=(-1)^{[n/2]}|\alpha|^2e^0\wedge\cdots\wedge e^9$ where $\varepsilon^{0123456789}=1$. We have
$$
*(X\wedge*\alpha)=(-1)^ni_X\alpha\,,\qquad
*d*\alpha=\frac{1}{(n-1)!}e^{b_{n-1}}\wedge\cdots\wedge e^{b_1}\nabla^a\alpha_{ab_1...b_{n-1}}\,.
$$
We also use the shorthand notation $i_a\alpha=i_{\partial_a}\alpha=\frac{1}{(n-1)!}e^{a_n}\wedge\cdots\wedge e^{a_2}\alpha_{aa_2\cdots a_n}$.
}
\begin{equation}
\nabla_{(a}K_{b)}=0\,,\qquad dX+i_KH=0\,,\qquad i_KX=0\,,
\label{eq:KX}
\end{equation}
the generalized Einstein equation
\begin{equation}
R_{ab}=
-2\nabla_{(a}X_{b)}
+\tfrac12\langle i_aH,i_bH\rangle
+\tfrac12\mathcal F^{(1)}_a\mathcal F^{(1)}_b
+\tfrac12\langle i_a\mathcal F^{(3)},i_b\mathcal F^{(3)}\rangle
+\tfrac14\langle i_a\mathcal F^{(5)},i_b\mathcal F^{(5)}\rangle
-\tfrac14(|\mathcal F^{(1)}|^2+|\mathcal F^{(3)}|^2)\eta_{ab}\,,
\label{eq:Einstein}
\end{equation}
the equations of motion for $B$ and $X$
\begin{align}
d*H+2X\wedge*H-2*dK-\mathcal F^{(1)}\wedge*\mathcal F^{(3)}+\mathcal F^{(3)}\wedge\mathcal F^{(5)}=&\,0\,,
\label{eq:B-eom}
\\
*d*X-2|X|^2-2|K|^2+\tfrac12|H|^2-\tfrac12|\mathcal F^{(3)}|^2-|\mathcal F^{(1)}|^2=&\,0\,,
\label{eq:X-eom}
\end{align}
and the generalized RR equations of motion
\begin{equation}
*d*\mathcal F^{(1)}-\langle X,\mathcal F^{(1)}\rangle-\langle H,\mathcal F^{(3)}\rangle=0\,,\qquad
*d*\mathcal F^{(3)}-i_X\mathcal F^{(3)}+K\wedge\mathcal F^{(1)}-*(H\wedge\mathcal F^{(5)})=0\,,
\end{equation}
with the five-form self-dual as usual, $*\mathcal F^{(5)}=\mathcal F^{(5)}$, and `Bianchi identities'
\begin{align}
i_K\mathcal F^{(1)}=&\,0\,,\\
d\mathcal F^{(1)}+X\wedge\mathcal F^{(1)}-i_K\mathcal F^{(3)}=&\,0\,,\\
d\mathcal F^{(3)}+X\wedge\mathcal F^{(3)}-H\wedge\mathcal F^{(1)}-i_K\mathcal F^{(5)}=&\,0\,,\\
d\mathcal F^{(5)}+X\wedge\mathcal F^{(5)}-H\wedge\mathcal F^{(3)}+*(K\wedge\mathcal F^{(3)})=&\,0\,.
\end{align}
The equations of motion and Bianchi identities for the generalized RR field strengths can be compactly encoded in a single equation for the anti-symmetric $32\times 32$ bispinor
\begin{equation}
\mathcal S^{\alpha i\beta j}=-\left(i\sigma^2\gamma^aF^{(1)}_a+\tfrac16\sigma^1\gamma^{abc}F^{(3)}_{abc}+\tfrac{1}{2\cdot5!}i\sigma^2\gamma^{abcde}F^{(5)}_{abcde}\right)^{\alpha i\beta j}
\end{equation}
as
\begin{equation}
\gamma^a\nabla_a\mathcal S-(X_a+\sigma^3 K_a)\gamma^a\mathcal S+\tfrac18H_{abc}\gamma^a\sigma^3\mathcal S\gamma^{bc}+\tfrac{1}{24}H_{abc}\gamma^{abc}\sigma^3\mathcal S=0\,.
\end{equation}

If $K$ vanishes the second equation in (\ref{eq:KX}) tells us that we can write $X=d\phi$ for some scalar field $\phi$.\footnote{This may not be true globally but we will only be interested here in local properties of the generalized supergravity equations.} It is then easy to see that the generalized supergravity equations reduce to the standard ones with $\phi$ being the dilaton. One can therefore think of these equations as a deformation of standard supergravity by the Killing vector field $K$.

It is interesting to ask whether there are solutions with $K\neq0$ which nevertheless solve also the standard supergravity equations. We will analyze the conditions for this to happen below. But first we will show that this indeed happens for the simplest class of solutions -- symmetric pp-wave backgrounds.

\section{Symmetric pp-wave solutions}
Symmetric space solutions are particularly simple to analyze since, by definition, the (gauge-invariant) supergravity fields (in the present case $X,H,\mathcal F^{(n)}$) must be proportional to invariant forms and therefore all terms involving derivatives of these in the supergravity equations drop out and we are left with a set of algebraic equations to solve. Of the symmetric spaces the simplest are pp-waves, or Cahen-Wallach spaces. A $d$-dimensional Cahen-Wallach space, $CW_d$, has metric
\begin{equation}
ds^2=2dx^+dx^-+A_{ij}x^ix^j(dx^-)^2+dx^idx^i\,,
\end{equation}
with $A_{ij}$ a non-degenerate symmetric bilinear form which can be taken to be diagonal $A=\mathrm{diag}(a_1,\ldots,a_{d-2})$. The only non-vanishing component of the Ricci tensor is $R_{--}=-\mathrm{tr}\,A$. The invariant forms are the constants (or volume form) together with $c_{i_1\cdots i_n} dx^-\wedge dx^{i_1}\wedge\cdots\wedge dx^{i_n}$ with $c_{i_1\cdots i_n}$ constant, e.g. \cite{Figueroa-OFarrill:2011tnp}.

We are then looking for solutions of the generalized supergravity equations of the form $CW_d\times\mathbbm{R}^{10-d}$. The analysis is very similar to the one performed in \cite{Figueroa-OFarrill:2012whx} for the standard type IIB case. Any invariant form consists of three pieces: the volume form on $CW_d$ wedged with some number of $dx^i$ ($i=d-1,...,8$) from $\mathbbm{R}^{10-d}$, a sum of other invariant forms on $CW_d$ wedged with invariant forms on $\mathbbm{R}^{10-d}$ i.e. something of the form $dx^-\wedge\cdots$ (without $dx^+$) and finally an invariant form on $\mathbbm{R}^{10-d}$. The norm of these are respectively negative, null and positive and we will write accordingly e.g. $\mathcal F^{(n)}=\mathcal F_-^{(n)}+\mathcal F_0^{(n)}+\mathcal F_+^{(n)}$. It is not hard to see that the generalized Einstein equation (\ref{eq:Einstein}) in the transverse $CW$-directions gives $H_-=\mathcal F_-^{(n)}=\mathcal F_+^{(n)}=0$ so that $|\mathcal F^{(n)}|^2=0$ (the LHS is zero and the RHS is a sum of negative terms). The components of the same equation in the Riemannian directions then imply that also $H_+=0$ so that all fluxes are null. Since $X$ is a one-form $X_-=0$ automatically and therefore $|X|^2\geq0$ and the $X$ equation of motion (\ref{eq:X-eom}) implies that $K$ is either time-like or null. However, if $K$ is time-like the remaining equations (e.g. $i_KH=0$) force $H=\mathcal F^{(n)}=0$ but this is inconsistent with the ($--$)-component of the generalized Einstein equation. We conclude that $K$ is null, which implies that also $X$ is null. From the remaining equations one finds $i_K\mathcal F^{(n)}=K\wedge\mathcal F^{(n)}=dK=i_KH=0$ but it is easy to see that this, together with the fact that $K$ is null, reduces the generalized supergravity equations to the standard ones (plus the decoupled Killing vector field $K$).

We have shown that any symmetric pp-wave solution of the generalized supergravity equations is in fact also a solution of the standard supergravity equations. This explains why this happened in \cite{Hoare:2016hwh,Roychowdhury:2018qsz}. Next we turn to the general question of under what conditions this happens.

\section{Trivial solutions}\label{sec:trivial}
We want to ask when solutions of the generalized supergravity equations are trivial, in the sense that they solve also the standard supergravity equations, even though $K\neq0$.

Assume first that $i_KH\neq0$. Then we see from (\ref{eq:KX}) that we cannot write $X=d\phi$ so there seems to be an obstruction to introducing the dilaton, which we need to make contact with standard supergravity. However, since all fields are isometric with respect to the isometry generated by $K$, it must be that $B$ transforms under the isometry by a gauge transformation, i.e. $\mathcal L_K B=d\Lambda_{(K)}$, where $\mathcal L_K=di_K+i_Kd$ is the Lie derivative along $K$, for some one-form $\Lambda_{(K)}$ which depends on $K$. If we cancel this by a compensating $B$-field gauge transformation we have $0=\mathcal L_K B=i_KH+di_KB$ and we can solve the equation for $dX$ in (\ref{eq:KX}) by taking
\begin{equation}
X=d\phi+i_KB\,.
\end{equation}
Now we observe that for $g,B,\phi$ to solve the standard supergravity equations it is of course necessary that $\phi$, the would-be dilaton, be invariant under gauge transformations of the $B$-field. This is not generically true when we solve for $X$ as above since $X$ is invariant (by definition) but $i_KB$ is not. Note that in solving for $X$ we have partially gauge fixed the $B$-field gauge invariance by requiring $\mathcal L_K B=0$. Therefore a necessary condition to get a standard supergravity solution is that the gauge transformation of $i_KB$ must vanish for all transformations preserving the gauge condition, i.e.
\begin{equation}
i_Kd\Lambda=0\,,\qquad\forall\Lambda\quad\mbox{such that}\quad di_Kd\Lambda=0\,.
\end{equation}
Next we note that we may take $\Lambda=K$ since $di_KdK=d\mathcal L_KK=0$ ($\mathcal L_KK=0$ is easily seen to follow from the fact that $K$ is a Killing vector). Therefore we find the condition $i_KdK=0$ or equivalently $|K|^2=$ constant (the integral curves of $K$ are therefore geodesics). In fact we can take a more general gauge parameter $\Lambda=fK$ where $f$ is any isometric function, $i_Kdf=0$. Then we find $i_Kd\Lambda=fi_KdK-i_K(df\wedge K)=-df|K|^2=-d(f|K|^2)$, and for this to vanish for general $f$ we must have that $K$ is null,
\begin{equation}
|K|^2=0\,.
\end{equation}
We will now argue that $i_KB\propto K$. If we assume that also the generalized RR field strengths $\mathcal F^{(n)}$ solve the standard supergravity equations it follows from the generalized Einstein equation (\ref{eq:Einstein}) that $i_KB$ is Killing and from the e.o.m. for $X$ (\ref{eq:X-eom}) that it must then also be null for this equation to reduce to the corresponding standard supergravity one (i.e. the same equation with $K=0$). Since $i_KB$ is also orthogonal to the null vector $K$ it must in fact be proportional to $K$, as claimed. It is perhaps not obvious that assuming that the generalized RR field strengths become directly the standard ones, without some $K$-dependent redefinitions, gives the most general possibility. However, we can still reach the same conclusion without this assumption as follows. Taking the trace of the generalized Einstein equation (\ref{eq:Einstein}) and adding twice the e.o.m. for $X$ (\ref{eq:X-eom}) we get an equation which does not involve the RR fields. Comparing to the corresponding equation in standard supergravity we get the condition
\begin{equation}
*d*X'+e^{2\phi}|X'|^2=0\,,
\end{equation}
where $X'=e^{-2\phi}i_KB$. Integrating this equation we find that the integral of the normal component of $X'$ over a surface equals the integral of $e^{2\phi}|X'|^2$ over the volume enclosed. This gives a complicated non-local expression for $X'$ which does not seem sensible, unless $|X'|^2=0$. Since $X'$ is null and orthogonal to the null vector $K$ it must again be proportional to $K$.

Using $X=d\phi+i_KB$ (note that this implies that $\phi$ is isometric, $i_Kd\phi=0$), $|K|^2=0$ and $i_KB=fK$, with $f$ an arbitrary isometric function (as follows from $d*X'=0$), the $X$ e.o.m. (\ref{eq:X-eom}) reduces to the standard equation of motion for the dilaton provided that $\frac12|\mathcal F^{(3)}|^2+|\mathcal F^{(1)}|^2$ reduces to the same expression in terms of RR field strengths $\mathcal F^{(n)}=e^\phi[dC^{(n)}+\ldots]$. Contracting the generalized Einstein equation (\ref{eq:Einstein}) with $K^b$ we find that also $2\langle i_a\mathcal F^{(3)},i_K\mathcal F^{(3)}\rangle+\langle i_a\mathcal F^{(5)},i_K\mathcal F^{(5)}\rangle+|\mathcal F^{(1)}|^2K_a$ must reduce to the same with $\mathcal F^{(n)}=e^\phi F^{(n)}$ for any value of the index $a$. From this it is clear that we must take the generalized RR field strengths to reduce to the standard RR field strengths (this argument does not rule out some very special exceptions of course). The remaining components of the generalized Einstein equation then forces $f$ to be a constant. The remaining equations now imply $f^2=1$ and since the sign of $K$ is irrelevant (a sign change of $K,H,\mathcal F^{(3)}$ leaves all equations invariant) we find
\begin{equation}
X=d\phi-K\,,\qquad |K|^2=0\,,\qquad dK=i_KH\,,\qquad K_a\gamma^a\mathcal S(1+\sigma^3)=0\,,
\label{eq:trivial-def1}
\end{equation}
where the last equation for the RR field strengths is equivalent to
\begin{equation}
i_K\mathcal F^{(2n+1)}=-K\wedge\mathcal F^{(2n-1)}\,,\qquad n=0,1,2\,.
\label{eq:trivial-def2}
\end{equation}
In this derivation we assumed that $i_KH\neq0$. Let's now look at the case when $i_KH=0$ so that $dX=0$ and we can directly write $X=d\phi$. Taking the trace of the generalized Einstein equation plus twice the $X$ e.o.m. again implies that $K$ is null. Proceeding as above we find also that $dK=0$ and the same conditions on the RR field strengths. We conclude that (except for possibly some very special cases) the conditions for a generalized supergravity solution to be also a standard supergravity solution is precisely (\ref{eq:trivial-def1}). Note that the case when $dK=0$, which is the case for the symmetric pp-waves discussed in the previous section, is somewhat trivial since in that case we can write $K=du$ and therefore $K$ can be removed by a shift of the dilaton $\phi\rightarrow\phi+u$ (an explicit example can be found in appendix B of \cite{Hoare:2016hwh} and, in hindsight, also the pp-wave background discussed in \cite{Hoare:2014pna}).

Note also that the condition $i_KB=-K$, which we found, can also be written $K^m(g_{mn}+B_{mn})=0$ so that $K$ is a null vector of the generalized metric $g+B$. The importance of this condition was noted in the example found very recently in \cite{Sakamoto:2018krs}, a generalization of a YB deformation of $AdS_3\times S^3\times T^4$ (to allow for non-zero $H$) with $R=J_{01}\wedge(p_0+p_1)$ (cf. \cite{Borsato:2016ose}), which unlike the pp-waves has $dK\neq0$.

We now turn to a derivation of the same conditions from the vanishing of the non-abelian T-duality anomaly in the case of YB sigma models.

\section{Anomaly for Yang-Baxter sigma models}\label{sec:YB}
As mentioned in the introduction there is some tension between the statement that we can have solutions with non-zero $K$ which also solve the standard supergravity equations, i.e. they should define one-loop Weyl invariant sigma models, and the expectation from non-abelian T-duality that when $K\neq0$, i.e. the algebra is non-unimodular, there should be an anomaly. Here we will resolve this tension by showing, on the example of bosonic homogeneous YB models, that in fact the anomaly goes away precisely when the conditions (\ref{eq:trivial-def1}) are satisfied. Note that YB models are a special case of so-called DTD models, obtained by adding a closed B-field and performing non-abelian T-duality (or T-dualizing on a centrally extended subalgebra) \cite{Hoare:2016wsk,Borsato:2016pas,Borsato:2017qsx}. We will work with this special case here since these models have been of some interest in the literature and since the expressions are somewhat simpler than the general non-abelian T-duality case. For these models we can also directly use the expressions for the target space fields derived in \cite{Borsato:2016ose}. Although, as we will see, the form of the background fields for these models does not actually allow for non-trivial solutions of (\ref{eq:trivial-def1}). It is clear however that the results will extend in a simple way to general DTD supercoset models and in particular to non-abelian T-duality of supercoset models since the calculations for these are essentially identical to the YB ones \cite{Borsato:2017qsx}. Note that in particular it should be straightforward to realize the examples found in \cite{Sakamoto:2018krs} as DTD models by starting from the $AdS_3\times S^3\times T^4$ supercoset with non-zero $H$. In fact the result should be valid even when one does not start from a supercoset model.

The general target space geometry for YB sigma models was derived in \cite{Borsato:2016ose}. Here we will only consider the bosonic case so we will set fermions and fermionic components of the R-matrix to zero (the geometry for this case can also be found in \cite{Kyono:2016jqy}). From appendix B of \cite{Borsato:2016ose} we have the expression for the Killing vector $K$,
\begin{equation}
K^a
=-\frac\eta2\widehat{\mathcal K}^{IJ}\tr\left([T_I,RT_J]\Ad_g(1+\Ad_h^{-1})P^a\right)
=
-(1+\Ad_h)^a{}_b[\eta R_g]^{bc}{}_c\,.
%
\end{equation}
Here $\eta$ is the deformation parameter appearing together with the anti-symmetric matrix $R$ defined on (a subalgebra of) the isometry algebra, e.g. $\mathfrak g=\mathfrak{so}(2,4)\times\mathfrak{so}(6)$, and satisfying the classical Yang-Baxter equation, $[RX,RY]-R([RX,Y]+[X,RY])=0$ $\forall X,Y\in\mathfrak g$. The generators of $\mathfrak g$ are denoted $T_I$ and $\widehat{\mathcal K}^{IJ}$ is the non-degenerate metric defined by the trace. The adjoint action by an element of the isometry group $G$ is $\Ad_gX=gXg^{-1}$ and $R_g=\Ad_g^{-1}R\Ad_g$. The element $h$ gives a local Lorentz-transformation and is defined in a certain way in terms of $g$ and $R$, see \cite{Borsato:2016ose} for further details. We can write $K$ in a way that will be more useful for our purposes as
\begin{equation}
K
=
-\eta n^I\tr\left(\Ad_g^{-1}T_IR_g[A_+^{(2)}+A_-^{(2)}]\right)
=
\frac12n^I\tr\left(\Ad_g^{-1}T_I[A_+-A_-]\right)
=
\frac12\tr\left(n_g[A_+-A_-]\right)\,,
\label{eq:K-YB}
\end{equation}
or, in components,
\begin{equation}
K_a=\eta[(1+\Ad_h)R_gn_g]_a\,.
\end{equation}
Here we have defined $n_I=\tilde f^J{}_{JI}$, the trace of the structure constants for the subalgebra where $R$ is defined (we have used the fact that $R^{IJ}f^K{}_{IJ}=2R^{KI}n_I$). Note that $n_I=0$ if this subalgebra is unimodular. Here we are interested in the non-unimodular case where $K$ is non-vanishing. We have also defined $n_g=\Ad_g^{-1}n$, $n=n_IT^I$. The one-forms $A_\pm=\mathcal O_\pm^{-1}(g^{-1}dg)$ where $\mathcal O_\pm=1\pm2\eta R_gP^{(2)}$ and $P^{(2)}$ denotes the projection on the translational (or "coset") generators $P_a$ of $\mathfrak g$ while $P^{(0)}$ projects on the Lorentz generators $J_{ab}$. We have $A_+^{(2)}=e^aP_a$ with $e^a$ the vielbein of the generalized supergravity background and $A_-^{(2)}=\Ad_h^{-1}A_+^{(2)}$ the Lorentz-rotated vielbein \cite{Borsato:2016ose}. We will need the square of $K$,
\begin{equation}
|K|^2
=
-\eta^2\tr\left(n_gR_gP^{(2)}(2+\Ad_h+\Ad_h^{-1})R_gn_g\right)\,.
\label{eq:K2-YB}
\end{equation}

We will also need the expression for $i_KB$. From \cite{Borsato:2016ose} we have $B_{ab}=2\eta[R_g]_{ab}$ and we get
\begin{align}
i_KB=&\,
e^bK^aB_{ab}
=
-2\eta^2e^b\,[R_gP^{(2)}(1+\Ad_h)R_gn_g]_b
=
-2\eta^2\tr\left(A_+^{(2)}R_gP^{(2)}(1+\Ad_h)R_gn_g\right)\nonumber\\
=&\,
2\eta^2\tr\left(P^{(2)}R_g(A_+^{(2)}+A_-^{(2)})R_gn_g\right)
=
\eta\tr\left(n_gR_g(A_+^{(2)}-A_-^{(2)})\right)\nonumber\\
=&\,
-\tfrac12\tr\left(n_g[A_++A_--2g^{-1}dg]\right)\,,
\label{eq:iKB-YB}
\end{align}
where we have used the definition of $A_\pm$.

We now turn to the question of the anomalous terms in the YB sigma model action. These follow from those for non-abelian T-duality by carrying out the field redefinition which relates these to the YB model \cite{Borsato:2016pas}. This field redefinition is complicated but it plays no role for the present discussion. The anomalous terms come from an extra term in the first order sigma model action which is the first step in non-abelian T-duality. This term is \cite{Elitzur:1994ri}, following the notation of \cite{Hoare:2016wsk},
\begin{equation}
L_{\mathrm{non-local}}=\alpha'\sigma n_I\partial^iA_i^I\,,
\end{equation}
and comes from the Jacobian for the change of variables $g^{-1}dg\rightarrow A$ in the path integral. Note that the conformal factor $\sigma=\partial^{-2}\sqrt{g}R^{(2)}$ is non-local in the worldsheet metric. Note also the $\alpha'$ signifying that this is a one-loop effect. It is now a simple matter to carry out the non-abelian T-duality (i.e. integrate out $A$), with this term included, and then the field redefinitions leading to the YB model following the steps in \cite{Borsato:2017qsx}. One finds that the YB model Lagrangian including the non-local anomaly terms takes the form (we drop the dilaton term since it is not needed for the present analysis)\footnote{For an attempt to derive (some of) the generalized supergravity equations by varying with respect to $\sigma$ see \cite{Hong:2018tlp}.}
\begin{align}
L
=&
-\eta(\gamma^{ij}+\varepsilon^{ij})\tr\left[\left(P^{(2)}(g^{-1}\partial_ig)+\alpha'n_g\partial_i\sigma\right)\mathcal O_+^{-1}R_g\left(P^{(2)}(g^{-1}\partial_jg)+\alpha'n_g\partial_j\sigma\right)\right]
\nonumber\\
&{}\quad+\tfrac12(\gamma^{ij}+\varepsilon^{ij})\tr\left[(g^{-1}\partial_ig)P^{(2)}(g^{-1}\partial_jg)\right]\,.
\label{eq:L-YB}
\end{align}
Here we have used the fact that $n$ is invariant under conjugation by an element of the subgroup where $R$ is defined. Let us look at the terms linear in $n$ first. Using $A_\pm=\mathcal O_\pm^{-1}(g^{-1}dg)$ and $\mathcal O_\pm=1\pm2\eta R_gP^{(2)}$ they can be written
\begin{equation}
L_n
=
\frac{\alpha'}{2}\gamma^{ij}\partial_i\sigma\,\tr\left[n_g(A_{+j}+A_{-j}-2g^{-1}\partial_jg)\right]
+\frac{\alpha'}{2}\varepsilon^{ij}\partial_i\sigma\,\tr\left[n_g(A_{+j}-A_{-j})\right]\,.
\end{equation}
Using the expression for $K$ and $i_KB$ in (\ref{eq:K-YB}) and (\ref{eq:iKB-YB}) this can be written, using for simplicity worldsheet form notation with pull-backs to the worldsheet being understood and dropping the overall factor of $\alpha'$, as
\begin{equation}
-d\sigma\wedge*i_KB
+d\sigma\wedge K
\sim
-d\sigma\wedge*(i_KB+K)
+\sigma d*K
+\sigma dK\,,
\end{equation}
where in the second step we added and subtracted the same term and dropped total derivatives. Now
\begin{equation}
d*K=\nabla*e^a\,K_a+*e^a\wedge e^b\,\nabla_bK_a\,,
\end{equation}
but the last term vanishes by the Killing vector equation. Finally we have, adding and subtracting $\sigma i_KH$,
\begin{equation}
L_n\sim
-d\sigma\wedge*(i_KB+K)
+\sigma(\nabla*e^a\,K_a+i_KH)
+\sigma(dK-i_KH)\,.
\label{eq:Ln}
\end{equation}
The second term is precisely $\sigma$ times the equation of motion of the string sigma model projected along $K$. It can therefore be removed by a, albeit non-local, field redefinition. In fact, forgetting about higher order terms in $\alpha'$, this field redefinition is simply an isometry shift $X^m\rightarrow X^m+\alpha'\sigma K^m$ (equivalently the term in question is proportional to the divergence of the isometry Noether current $J=K-*i_KB$, e.g. \cite{Wulff:2014kja}). One might worry that such a shift, being non-local, would not be allowed. However, the non-locality here is only in the worlsheet metric and not in the dynamical fields $X^m$ themselves and furthermore, being a simple shift, this change of variables does not lead to a non-trivial Jacobian from the path integral measure. Another justification for dropping these terms is that, given our earlier analysis of the generalized supergravity equations, this leads to a sigma model whose target space solves the standard supergravity equations and which is therefore Weyl invariant and non-anomalous. The remaining terms in (\ref{eq:Ln}) vanish precisely when the conditions (\ref{eq:trivial-def1}) are satisfied which is what we wanted to show.

Although they are higher order in $\alpha'$ let us also consider the $n^2$-terms in (\ref{eq:L-YB}). They are
\begin{align}
L_{n^2}=&\,
-\alpha'^2\eta\gamma^{ij}\partial_i\sigma\partial_j\sigma\,\tr\left(n_g\mathcal O_+^{-1}R_gn_g\right)
=
-\alpha'^2\eta\gamma^{ij}\partial_i\sigma\partial_j\sigma\,\tr\left(n_g[\mathcal O_+^{-1}-1]R_gn_g\right)\nonumber\\
=&\,
2\alpha'^2\eta^2\gamma^{ij}\partial_i\sigma\partial_j\sigma\,\tr\left(n_gR_gP^{(2)}\mathcal O_+^{-1}R_gn_g\right)
=
\alpha'^2\eta^2\gamma^{ij}\partial_i\sigma\partial_j\sigma\,\tr\left(n_gR_gP^{(2)}[1+\Ad_h]R_gn_g\right)\nonumber\\
=&\,-\tfrac12\alpha'^2\gamma^{ij}\partial_i\sigma\partial_j\sigma\,|K|^2\,,
\end{align}
where in the last step we used (\ref{eq:K2-YB}). Again this vanishes precisely when $K$ is null in accordance with (\ref{eq:trivial-def1}). This resolves the apparent tension between the fact that we have a supergravity solution with $K\neq0$, which should be non-anomalous, on the one hand, and the expectation from non-abelian T-duality that $K\neq0$ should imply an anomaly on the other, by showing that this expectation is too naive and there can be special cases where the anomalous terms cancel in a non-trivial way.

The $\sigma$-dependent terms take a simple geometrical form when expressed in terms of the generalized supergravity fields $K$ and $X$, namely (now we include the standard dilaton term)
\begin{equation}
L_\sigma=
\alpha'd\sigma\wedge J
-\alpha' d*d\sigma\,\phi
-\tfrac12\alpha'^2d\sigma\wedge*d\sigma\,|K|^2
=
\alpha'd\sigma\wedge K
-\alpha'd\sigma\wedge*X
-\tfrac12\alpha'^2d\sigma\wedge*d\sigma\,|K|^2\,,
\end{equation}
where $J=K-*i_KB$ is the isometry Noether current and we have dropped a total derivative in the last equality. We expect this simple form for the anomalous terms, the generalized supergravity analog of the standard dilaton term, to be valid generally.\footnote{Recall from our previous discussion that the dilaton term by itself is not gauge-invariant in generalized supergravity and must be completed to $X$. This shows that the second term should be there in general. Then we can argue as above that the first term must also be there so that the anomalous terms cancel for trivial solutions of the generalized supergravity equations.}

Unfortunately, for the standard (bosonic) YB sigma models considered here, obtained by starting with a coset sigma model without WZ term, it is not hard to see that the condition $K+i_KB=0$ forces $[n_gR_g]_a=0$ which actually implies that $K$ vanishes. Therefore the Weyl invariant YB models are precisely the unimodular ones of \cite{Borsato:2016ose}. The anomaly analysis here should however apply more generally, with only minor modifications, to all models constructed using non-abelian T-duality, in particular, as already mentioned, to the examples of \cite{Sakamoto:2018krs} where $K$ is not forced to vanish. It would be interesting to find in which classes of models one can avoid the anomaly in this way.

\section{Conclusion}
We have shown that the generalized supergravity equations, which follow from kappa symmetry of the Green-Schwarz superstring (or one-loop scale invariance), can have `trivial' solutions in the sense that they solve also the standard supergravity equations. We have argued that this happens precisely when the conditions in (\ref{eq:trivial-def1}) are satisfied, in particular the Killing vector $K$ should be null. All symmetric pp-wave solutions are in fact of this type as we have seen.\footnote{One could also obtain such solutions by starting with a generalized supergravity solution and taking a boosted limit where $K$ becomes null, e.g. \cite{Sfetsos:1994fc}. I thank A. Tseytlin for this comment.} The tension with the expectation from non-abelian T-duality that $K\neq0$, which corresponds to non-abelian T-duality on a non-unimodular algebra, should be anomalous was resolved, in the specific context of bosonic YB models, by showing that in fact, upon a non-local field redefinition, the anomalous terms cancel for these backgrounds. It would be nice to find an interpretation for this non-local field redefinition.

In the case of standard (bosonic) YB models, one finds that the remaining conditions on $K$ do not have any non-trivial solutions and therefore this class of models does not seem to realize the possibility of canceling the anomaly in a non-trivial way. The extension to general non-abelian T-duality, or more generally DTD models \cite{Borsato:2016pas}, and in particular the interesting examples of \cite{Sakamoto:2018krs}, should be straight-forward however and one should be able to recover also the conditions on the RR fields (\ref{eq:trivial-def2}) from the analysis of the anomaly. Perhaps this analysis can also be extended to the case of Poisson-Lie T-duality, cf. \cite{Hoare:2017ukq,Lust:2018jsx}. It would be interesting to find what is needed in order to have non-trivial solutions of this form.

Another related point, which we have not addressed so far, is what happens to the (local) terms in the sigma model action that depend on $K$. These appear first at the quartic order in fermions, as is easily seen from \cite{Wulff:2013kga}, in light of the generalized supergravity constraints in \cite{Wulff:2016tju}. In fact since $K$ is a null isometry it is natural to use a kappa symmetry gauge fixing adapted to this isometry, $K_a\gamma^a\theta=0$, and in this gauge it is not hard to show that the terms involving $K$ go away leaving us with the standard Green-Schwarz action. Even without this kappa symmetry gauge fixing our analysis guarantees that it must be possible to remove these terms by a field redefinition in the non-anomalous cases.

We leave the question of whether there is any deeper significance to these backgrounds for the future.

\vspace{1cm}

\section*{Acknowledgments}
It is a pleasure to thank Riccardo Borsato, Ben Hoare and Stijn van Tongeren for interesting discussions related to this topic, and Arkady Tseytlin for useful discussions and helpful comments on a draft of this note.

\vspace{0.5cm}

\bibliographystyle{nb}
\bibliography{biblio}{}

\end{document}